\documentclass[letterpaper, 10 pt, conference]{ieeeconf}
\IEEEoverridecommandlockouts
\usepackage{graphicx} 
\usepackage{epsfig} 
\usepackage{mathptmx} 
\usepackage{times} 

\newcommand{\I}{\hspace*{1cm}}
\newenvironment{algo}{\begin{tt}\ \\ }{\end{tt}}
\newtheorem{definition}{Definition}[section]
\newtheorem{theorem}{Theorem}[section]

\title{\LARGE \bf Abstraction and Refinement in Static Model-Checking }
 
\author{Kaninda Musumbu \\
LaBRI~ (UMR 5800 du CNRS),\\
Universit\'e Bordeaux 1, France \\
351, cours de la Lib\'eration, F-33.405 TALENCE Cedex\\
e-mail: musumbu@labri.fr}

\begin{document}

\maketitle
\thispagestyle{empty}
\pagestyle{empty}

\begin{abstract}
Abstract interpretation is a general methodology for building static
analyses of programs.  It was introduced by P. and R. Cousot in \cite{cc}.
We present, in this paper, an application of a generic abstract interpretation 
to domain of model-checking.
Dynamic checking are usually easier to use, because the concept are established
 and wide well know. But they are usually limited to systems whose states 
space is finite. In an other part, certain faults
cannot be detected dynamically, even by keeping track of the history of the 
states space.Indeed, the classical problem of finding the right test cases is 
far from trivial and limit the abilities of dynamic checkers further.
Static checking have the advantage that they work on a more abstract level than
 dynamic checker and can verify system properties for all inputs. Problem, it
 is  hard to guarantee that a violation of a modeled property corresponds to a 
fault in the concrete system. We propose an approach, in which we
generate  counter-examples dynamically using the abstract interpretation
techniques.
  
\end{abstract}


\section{Introduction}
Being given that the number of state of a model believes in an exponential way 
with the number of variables and components of the system, the model-checking 
became complicated to treat in an automatic way. In order to make this work 
realizable, it is necessary to reduce the sizes of these models with an aim of 
reaching time and reasonable memory capacities. The techniques of reduction 
seek to suppress the harmful effects of the combative explosion. When the 
graphs of behavior comprise several million or milliards of states and 
transitions, the physical limits of the memory are quickly reached. 
It is then necessary to resort to techniques compressions of the graphs of 
behavior. Most known is based on the BDD (Binary Decision Diagrams). 
At the enumeration time, to decide if a reached state was already met requires 
to traverse the explored part of the graph. This subgraph, which does not 
cease growing bigger, must be arranged in the read-write memory. The limits of 
this memory are  quickly exceeded and the implementation of algorithms of pagination know  a considerable fall of performances. The methods of abstraction make
 it possible to eliminate the proliferation from different states(ones from the other) by possibly unimportant details within sight of the properties to be 
checked. It is essential that the small-scale model preserve sufficient 
information to produce the same results as the models of origin and to 
preserve the same properties that one wishes to check. These two exigences 
must be considered with attention at the time of the generation of a abstract 
model  starting from a concrete model. To conceive a ``good''  method of 
reduction consists to produce  a reduction relation verifying three criteria:
 an important reduction ratio, a relation of strong preservation and an easy 
deduction of the relation of reduction starting from the description of the 
system, the ideal being the construction of the reduced graph directly starting
 from the description. The way whose details of the abstraction will be 
selected for the checking can be made in an automatic or manual way. 
The manual technique  includes abstract interpretations selected by the user. 
The abstractions considered generally preserve the properties in a weak way, 
which means that they are only preserved abstracted model  with the concrete 
model. Thus, if one can guarantee that a property is checked, that  is  
different with  its negation. 
The abstract interpretation is a methodology aiming at defining, analyzes and justifier your techniques of approximate computation  of properties of systems 
in \cite{cc}.
Whatever the semantics may be used. It then consists in placing 
the analysis not in the concrete domain  but in a  abstract domain, 
(simplified and limited) which conserves the search properties, the major 
disadvantage is that the results are in general less precise and that one 
needs accommodate approximations of the properties. In our paper, we present a 
technique of abstraction called {\em abstraction by  predicate} 
of a refinement to reduce the generality and the minimality of the analysis, thus a violation of a property detected on one of  abstract path has a strong probability of existing on a path of the concrete model.
Analysis is made at the global state space level: traversal algorithm (similar to the one used to build the state space) is used to check out deadlock, 
livelock or divergent states. Example pathes starting
from the initial state and leading to a deadlock, livelock or divergent state can be extracted. To this
end, we have to collect during the search of these special states the intermediate state sets reached
before them. verication is based on bisimulation minimizations and comparisons .

\section{Generality}

\begin{center} {\bf Small Example: 
The rule of Signs} \mbox{~~~~~~~} \\ \mbox{~~~~~~~} \\ 
\mbox{~~~~~~~} \\ 
\centerline{\setlength{\unitlength}{0.012500in}%
\begingroup\makeatletter
\def\x#1#2#3#4#5#6#7\relax{\def\x{#1#2#3#4#5#6}}%
\expandafter\x\fmtname xxxxxx\relax \def\y{splain}%
\ifx\x\y   
\gdef\SetFigFont#1#2#3{%
  \ifnum #1<17\tiny\else \ifnum #1<20\small\else
  \ifnum #1<24\normalsize\else \ifnum #1<29\large\else
  \ifnum #1<34\Large\else \ifnum #1<41\LARGE\else
     \huge\fi\fi\fi\fi\fi\fi
  \csname #3\endcsname}%
\else
\gdef\SetFigFont#1#2#3{\begingroup
  \count@#1\relax \ifnum 25<\count@\count@25\fi
  \def\x{\endgroup\@setsize\SetFigFont{#2pt}}%
  \expandafter\x
    \csname \romannumeral\the\count@ pt\expandafter\endcsname
    \csname @\romannumeral\the\count@ pt\endcsname
  \csname #3\endcsname}%
\fi
\endgroup
\begin{picture}(120,104)(60,716)
\thicklines
\put( 60,716){\framebox(120,104){}}
\put( 60,794){\line( 1, 0){120}}
\put(180,794){\line( 0, 1){  0}}
\put( 90,820){\line( 0,-1){104}}
\put(101,806){\makebox(0,0)[lb]{\smash{\SetFigFont{9}{10.8}{rm}
$+$
}}}
\put(131,806){\makebox(0,0)[lb]{\smash{\SetFigFont{9}{10.8}{rm}
$0$
}}}
\put(161,806){\makebox(0,0)[lb]{\smash{\SetFigFont{9}{10.8}{rm}
$-$
}}}
\put( 71,754){\makebox(0,0)[lb]{\smash{\SetFigFont{9}{10.8}{rm}
$0$
}}}
\put( 71,730){\makebox(0,0)[lb]{\smash{\SetFigFont{9}{10.8}{rm}
$-$
}}}
\put( 71,776){\makebox(0,0)[lb]{\smash{\SetFigFont{9}{10.8}{rm}
$+$
}}}
\put(101,776){\makebox(0,0)[lb]{\smash{\SetFigFont{9}{10.8}{rm}
$+$
}}}
\put(131,776){\makebox(0,0)[lb]{\smash{\SetFigFont{9}{10.8}{rm}
$0$
}}}
\put(158,776){\makebox(0,0)[lb]{\smash{\SetFigFont{9}{10.8}{rm}
$-$
}}}
\put(101,754){\makebox(0,0)[lb]{\smash{\SetFigFont{9}{10.8}{rm}
$0$
}}}
\put(131,754){\makebox(0,0)[lb]{\smash{\SetFigFont{9}{10.8}{rm}
$0$
}}}
\put(158,754){\makebox(0,0)[lb]{\smash{\SetFigFont{9}{10.8}{rm}
$0$
}}}
\put(101,730){\makebox(0,0)[lb]{\smash{\SetFigFont{9}{10.8}{rm}
$-$
}}}
\put(131,730){\makebox(0,0)[lb]{\smash{\SetFigFont{9}{10.8}{rm}
$0$
}}}
\put(158,730){\makebox(0,0)[lb]{\smash{\SetFigFont{9}{10.8}{rm}
$+$
}}}
\put( 68,806){\makebox(0,0)[lb]{\smash{\SetFigFont{9}{10.8}{rm}
$\stackrel{\sim}{\times}$
}}}
\end{picture}
} 
{Such that the following diagram commutates} \\ 
\mbox{~~~~~~~} \\ 
\mbox{~~~~~~~} \\ 
\centerline{\setlength{\unitlength}{0.012500in}%
\begingroup\makeatletter
\def\x#1#2#3#4#5#6#7\relax{\def\x{#1#2#3#4#5#6}}%
\expandafter\x\fmtname xxxxxx\relax \def\y{splain}%
\ifx\x\y   
\gdef\SetFigFont#1#2#3{%
  \ifnum #1<17\tiny\else \ifnum #1<20\small\else
  \ifnum #1<24\normalsize\else \ifnum #1<29\large\else
  \ifnum #1<34\Large\else \ifnum #1<41\LARGE\else
     \huge\fi\fi\fi\fi\fi\fi
  \csname #3\endcsname}%
\else
\gdef\SetFigFont#1#2#3{\begingroup
  \count@#1\relax \ifnum 25<\count@\count@25\fi
  \def\x{\endgroup\@setsize\SetFigFont{#2pt}}%
  \expandafter\x
    \csname \romannumeral\the\count@ pt\expandafter\endcsname
    \csname @\romannumeral\the\count@ pt\endcsname
  \csname #3\endcsname}%
\fi
\endgroup
\begin{picture}(195,125)(65,700)
\thicklines
\put(130,810){\vector( 1, 0){105}}
\put(135,705){\vector( 1, 0){105}}
\put(260,800){\vector( 0,-1){ 80}}
\put(100,720){\vector( 0, 1){ 80}}
\put(240,805){\makebox(0,0)[lb]{\smash{\SetFigFont{11}{13.2}{rm}$Sgn$}}}
\put( 80,700){\makebox(0,0)[lb]{\smash{\SetFigFont{11}{13.2}{rm}$Int \times Int$}}}
\put(245,700){\makebox(0,0)[lb]{\smash{\SetFigFont{11}{13.2}{rm}$Int$}}}
\put(260,755){\makebox(0,0)[lb]{\smash{\SetFigFont{11}{13.2}{rm}
$\gamma$
}}}
\put(170,710){\makebox(0,0)[lb]{\smash{\SetFigFont{11}{13.2}{rm}
$\times$
}}}
\put( 85,760){\makebox(0,0)[lb]{\smash{\SetFigFont{11}{13.2}{rm}
$\alpha$
}}}
\put(170,815){\makebox(0,0)[lb]{\smash{\SetFigFont{11}{13.2}{rm}
$\stackrel{\sim}{\times}$
}}}
\put( 65,805){\makebox(0,0)[lb]{\smash{\SetFigFont{11}{13.2}{rm}$Sgn \times Sgn$}}}
\end{picture}
} 
\end{center} 
{\bf Consistency (soundness):} 
{$$\forall x,y \in Int: X \times y \in \gamma(\alpha(x) 
\stackrel{\sim}{\times} \alpha(y)) $$} 

\subsection{Definition}
Abstract Interpretation is a general methodologies for
 automatic analysis of the run-time properties of system.. The problem is  that the exact analysis may be very  expensive, sometimes  through  decidable 
 properties may be NP-complete. The idea is to find a decidable  approximation 
which is soundness and calculable. 

\subsection{Mathematical theory of Abstract Interpretation}
 Often, AI refers to the concept of {\em connection galoisienne} a 4-tuple 
$(C, A, \alpha, \gamma)$ where  $C$
and $A$ are complete lattices ,  
$\gamma: A \rightarrow C$ and $\alpha: C 
\rightarrow A$ are monotonous functions \\ 
such as: $$
\begin{array}{lll} \forall \beta \in a: & \alpha(\gamma(\beta)) & = \beta, \\
 \forall C \in C: & \gamma(\alpha(c)) & \geq C. \\ 
\end{array}$$ \mbox{~~~~~~~~~} \\ 
Impossible into practice of {\it generating and analyzing} all possible traces 
of execution for a given program . 

\subsection{Motivation}
 Abstract interpretation is based on three fundamental ideas : \\ 
{\it abstract domain, abstract operators and point fixes computation} 
Abstract domain  and abstract operators are used to carry out a program on 
abstract values. Computation  of the fixed point: directs the process on 
abstract values (define in a certain way the semantics of the program) 
Objective: To obtain information on the execution and the results of program. 
Provided that the abstract domain and  operators  satisfy certain 
constraints.

\subsection{Semantics}
Its  definition has two view points: 
\begin{itemize} 
\item {\bf Theoretical} associates a meaning to objects handled by the 
programs.
 \item {\bf Piratical} associates a program a semantic function ( stomata). 
$$ < { \mathcal{P},e>\stackrel{\tau} {\longrightarrow} < \mathcal{P'},e'> } $$ 
$$\mathcal{P} = \tau.\mathcal{P'} \mbox{ and } e' =\tau(e)$$ 
$\tau$ is a transition related. Note:

 $$\tau[\mathcal{P}](e_0)=cal$$ if $cal=<\_,e_n>$  then the results of the 
program are the values of variables in the last state. 
\item {\bf Denotational} $$\tau:(D\longrightarrow D) \longrightarrow 
(D\longrightarrow D) $$ 
$$\tau=\lambda f.\lambda x.(\mbox{ if }  p(x) \mbox{ then } x \mbox{ else } f(h(x))\mbox{ fi}  $$
where $p$ is a predicate , h any function. 
Example F91McCarthy:
$$\tau=\lambda f.\lambda x.(\mbox{ if }  x> 100 \mbox{ then } x-10 \mbox{ else } f(f(x+11) \mbox{ fi}  $$
 \end{itemize}

\subsubsection{Abstract Domain} :
 Any program $P$ handles data which belong to a $D_{s}$ domain says standard. 
To make abstract interpretation will consist in choosing an abstraction of 
data $D_{abs}$ Ò
{First Approach} \\ 
$$\theta=\{x_1 \leftarrow t_1, \ldots, x_n \leftarrow t_n \} $$
\centerline{$\beta$ approximates $\theta$ iff} $$\beta=\{x_1 \leftarrow prop(t_1), 
\ldots, x_n \leftarrow prop(t_n) \} $$ this define  concretes semantics \\ 
More often  no-calculable. \\ 
{Construction process} $P$ can be consider like a  partial function 
 of $$ P:D_{s}^m \longrightarrow D_{s}^n, n,m 0 $$ 
 Example: function of McCarty known as function91 $$
F91(x) = \mbox{ if } x > 100 \mbox{ then } x-10 \mbox{ else } F91(F91(x+11))$$ 
 
\begin{algo} 
int F91(int x){\\
\I int F; \\
\I \I if (x >100) f =x-10;\\ 
\I \I else f=F95(F95(x+11));\
\I   return F; \\
} 
\end{algo} 
\begin{verbatim}
int F91McCarthy(void){
   int x; 
   scanf(&x); 
   printf("value of F91 of %d = %d ",
           x, F95(x)); 
   exit(0); 
} 
\end{verbatim}
{\bf Note::} There is not a proof of termination of 
$F91McCarthy$, $\forall X \in \mathcal{Z}$ \\ 
The idea is to replace $Z$ by its power set  $\cal{P}(Z)$. \\  
 We get the following  definition: \\
 
$\begin{array}{l}
  F91(X)=\{x-10: x > 100  \wedge x \in X \subseteq Z \} \cup\\ 
\mbox{~~} F91(F91(\{x+11: x \leq 100 \wedge  x \in X \subseteq Z \}))\\
\end{array}$\\

It is easy to show that 
$F9 1(C_1)=C_2$ verifies the condition: $\forall x \in C_1 \exists y 
\in C_2: y = f(x)$ \\

{\bf Note::} the calculation of  such function is 
too expensive for simple value, the definition of the operations on a such 
 domain is too complex. \\

{Second Approach} \\ To choose "a good" system of 
representation of properties 
 $$\beta=\{ x_1 \leftarrow \alpha(prop(t_1)), \ldots, x_n \leftarrow 
\alpha(prop(t_n)) \} $$ 
 Choice of an (judicious) approximation of each element of $\cal{P}(Z)$
by  an interval 
$[min..max]$  
$$D_{abs}=\{ [ s..t ]: s,t \in Z \cup \{-\infty, +\infty \} \} $$ \\ 
we define an order on $D_{abs}$, noted $\subseteq$ : 
$[s..t ] \subseteq [s'.t' ]$ iff $s\geq s' $ $t\leq t'$

\paragraph{ Lemma:} 
$(D_{abs}, \subseteq) $, is a lattice whose lower bound is $ [ ] $ and the 
upper bound  is $[-\infty, +\infty ] $. Abstraction and 
concretization function: 
$$\alpha: C \longrightarrow A: C \rightarrow \alpha(c)=[min(c)..max(c)]$$ 
$$\gamma: A  \longrightarrow C: a \rightarrow \gamma(a)=[s,s+1...,t-1,t]$$ 
with $a=[s..t]$ such that they verifying the constraints of coherence: 
$$ \forall c \in C: \gamma(\alpha(c)) \supseteq c \\
\forall a \in A: \alpha(\gamma(a)) = a.  $$    
\paragraph{ Remark}
\begin{itemize} 
\item an  equivalent abstract  of a program carries out the same standard  
operations that the original except that the domains are different. 
\item for a real  Pascal, C or Java programs,  the work of rewrite would be 
too tiresome. In fact one defines abstracted operators, the abstract interpretor
  uses those to carry out calculations on the abstract data by interpreting 
the program to be analyzed. 
\item In practice each operator or function of the language must have an 
abstract equivalent. 
The quality required is their consistency, their coherency with respect to their
 equivalent concrete operator. For the reason of performance, one requires 
the efficiency and convergence to guarantee a termination and acceptable 
computing time. 
\end{itemize} 
Abstract version of the F91 function: \\\\
$\begin{array}{ll}
F_{a}91([s..t]) = & [max(91, s-10)..(t-10) ] \cup \\
      &    F_{a}91(F_{a}91([(s+11)..min(t+11,111)]))
\end{array}$ 

$\begin{array}{ll}
\forall I_i, I_j \in D_a: & I_i=[s..t ], I_j=[s'..t' ] \Rightarrow \\
& I_i \cup I_j=lub(I_i, I_j) = [ min(s)..max(t, y ou) ], 
\end{array} $
The abstract calculus: 
$$F_{a}91([-\infty..+\infty])=[91, +\infty
 ] \cup \\ F_{a}91(F_{a}91([-\infty.., 111]))$$ 
$$F_{a}91([-\infty..111])=[91,101 ] \cup \\
 F_{a}91(F_{a}91([-\infty.., 111]))$$

{\bf Note::} The set of functions of $D_a \longrightarrow D_a$ can be 
provided with an order $ f\leq G $ iff $\forall I \in D_a: f(I)\subseteq 
g(I) $ \\ 
The fixpoints calculus:  it is useful at the time of the recursive calls, 
to ensure the termination while proceeding by successive 
approximations. \\

\paragraph{Complete lattice} 
\begin{itemize} 
\item  a lattice iff $\exists \bot \in D$ and $\exists \top \in D$ 
\item complete iff \begin{itemize} 
\item $\forall X \subseteq D \exists U \in D: \forall X \in X x\leq U $ and 
\item $\forall X \subseteq D \exists L \in D: \forall X \in X x\geq L $ 
\end{itemize} 
\end{itemize} 
It is obvious thats $(D, \leq)$ satisfy this conditions.  
    
\paragraph{ Monotonicity and continuity} 
Let  $A$ be a  complete lattice  with a partial order $\leq$ and 
$T :A \longrightarrow A$ a transformation 
\begin{itemize} 
\item $T$ is {\bf monotonous} iff $\forall X, y \in a: X \leq y 
\Rightarrow T(x) \leq T(y)$ 
\item $T$ is {\bf continuous} iff $\forall X \subseteq a: T(lub(X))=lub(T(X))$
 \end{itemize} 
The transformation that we consider is a functional from  a set of function in i
ts self. 
$$
T: (D_{a} \longrightarrow D_a) \longrightarrow (D_{a} \longrightarrow D_{a}) $$
   
$
\begin{array}{l}
(TF_{a}91)([s..t])=[max(91, s-10)..(t-10) ] \cup\\
 F_{a}91(F_{a}91([(s+11)..min(t+11,111)]))
\end{array}$

\paragraph{Lemma:}
 T is continuous and monotonous: 
\begin{itemize} \item $\forall I_i, I_j \in D_{a}: I_i \leq I_j 
\Rightarrow f(I_i) \leq f(I_j)$ \item $ \forall I_1\subseteq I_2\subseteq... 
\subseteq I_n \subseteq... \Rightarrow f(\cup_{i=1... \infty}I_i) = \cup_{i=1..
. \infty} f(I_i) $ 
\end{itemize}    
{Theorem} Let $f([s..t])$, the computing  fixpoint consist of $T$ to 
$f([s..t])$. 
\begin{itemize} 
\item If the constraints on the domain and the operators are satisfied: \\
 then any fixpoint of $T$ is a correct approximation of the function $f$ 
\item the smallest fixpoint of $T$ exists and constitutes the best approximation
 of $f$ 
\item the smallest fixpoint of $T$ coincide with the limit of an increasing 
\item the smallest fixpoint of $T$ coincide with the limit of an increasing 
sequence of approximation: $f_0 \leq f_1 \leq f_2 \leq f_0... \leq f_n \leq...$
\\ such as: \ \ $f_0(I)=\bot ~ \forall I \in D_a$ \\ 
$f_{k+1}=T(f_k>) ~ \forall k\geq 0$ 
\end{itemize} 

\subsubsection{Fixpoint Approach}
Fixpoint Approach is based on the monotonicity (continuity) of the 
transformation of the tuples set representing the {\tt pre} and {\tt post} 
condition for all predicate. 
Termination of the algorithm in the case of an infinite abstract domain, it 
did not guarantee. Which is the case if one makes an infinity  different  
recursive call. One can limit oneself to abstract fields finished in certain 
cases 
that can averrer genant itself or unacceptable. 
A possible solution, would be to replace an infinite sequence
 of approximation by a number of the approximate values.    

\paragraph{Approach Widening/Narrowing}

Suppose the abstract semantics of the program
 given by a function $f_P: D_{abs}\longrightarrow D_{abs}$. The analysis 
proceeds as follows: 
\begin{enumerate} 
\item {\em Widening}: calculation of sequences limit $X$ built by: 
\begin{itemize}
\item[] 
$x_0 = \bot$ 
\item[] $x_{i+1} = x_i \mbox{ and } f_P(x_i)) \sqsubseteq x_i $
\item[] \mbox{~~~~~~~~~ else}$ x_i \bigtriangledown f_P(x_i)$ 
\item {\em Narrowing} to improve the result obtained by the widening: by 
\end{itemize} 
calculating the sequences limit of  $Y$ built by:  
\begin{itemize}
\item[] $y_0 = \sqcup X$ 
\item[] $y_{i+1} = \mbox{and} f_P(y_i)=y_i \mbox{then} y_i$  
\item[] \mbox{~~~~~~~~~~ else} $ y_i \bigtriangleup f_P(y_i))$ 
\end{itemize} 
\end{enumerate}    
\paragraph{Properties} 
\begin{enumerate} 
\item {\em Widening}: 
$\bigtriangledown:l \times L \longrightarrow L$ \ \ $\forall X,Y \in L: ~~X 
\sqsubseteq Y \bigtriangledown Y $ and $ Y \sqsubseteq X \bigtriangledown Y $
 \\ $\rightarrow X \sqcup Y \subseteq X \bigtriangledown Y $ \\ 
$\bot \bigtriangledown X = X \bigtriangledown \bot = X $ 
\item {\em Narrowing} 
$\bigtriangleup:l \times L \longrightarrow L$ \\ 
$\forall X,Y \in L: ~~Y \sqsubseteq X \Rightarrow Y \subseteq X \bigtriangleup Y \sqsubseteq X$ \\ 
\end{enumerate}    
\paragraph{Widening applied to the intervals}
 $$\begin{array}{ll} [ l_0, u_0]
\bigtriangledown [ l_1, u_1]=&[ \mbox{ and } l_1 u_0 \mbox{then} +\infty \mbox{else} u_0 ] \\ 
\end{array}$$ 
Example instead of making the recursive call with $F_a91([-\infty..111]) $ 
one will do it with $F_a91([-\infty..+\infty])$. 
But a loss of precision would be introduced. This will allow to speed up the 
computation  of the fixpoint.

\section{Refinement}

\paragraph{Motivation} The abstract interpretation framework establishes a methodology based on rigorous semantics for constructing abstraction that overapproximate the behavior of the program, so that every behavior in the program is covered by a corresponding abstract execution. Thus, the abstract behavior can be exhautively checked for an invatiant in temporal logic. 
Refinement guided by counterexample consist on approximation of the set of sates that lie on a path from initial state to a bad state which is successsively refine that is done by forward or backward passes. This process is repeated until the fixpoint is reached. If the the resulting set of state is empty then the property is proven. Otherwise, yhe methode does not guaranties that the contreexample trace is genuine.

\subsection{Preliminaries}

\begin{definition}

\end{definition}
\begin{theorem} {\tt Cousot77}\\
Let $S=(Q,Q_{init},\sum,\rightarrow)$
 a system representing the semantics of  program. The system
$S^A=(Q^A,Q_{init}^A,\sum,\rightarrow^A)$
 is an abstraction of $S$ $\iff$ there exists a Galois connexion:\\
$ \alpha: \mathcal{P}(Q)\longmapsto  \mathcal{P}(Q^A),
\gamma: \mathcal{P}(Q^A)\longmapsto  \mathcal{P}(Q)$\\
such that
\begin{itemize}
\item $ Q_{init} \subseteq \gamma(Q_{init}^A) $
\item $ \forall \tau \in \sum, \forall Q_i^A \subseteq Q^A.
post[\stackrel{\tau}{\rightarrow}](\gamma(Q_i^A)) \subseteq
\gamma(post[\stackrel{\tau^A}{\rightarrow}](\gamma(Q_i^A)) $
\end{itemize}
\end{theorem}

\begin{definition}
{\bf Predicat Abstraction {\tt Graf\&Saidi97}} 
\paragraph{ Abstract State}

Let  $Prog=(\mathcal{V,T}=\{\tau_1,\ldots,\tau_n \}, Init)$ and
$\varphi_1,\ldots,\varphi_k$  predicates over the $Prog$'s variables\\
we define an abstraction
$S^A=(Q^A,Q_{init}^A,\sum,\rightarrow^A)$ as following:
\begin{itemize}
\item $Q^A = B^k$ , $Q^A$ is the valuations' set of  $k$ boolean variables, \\
any subset $Q^A$ can be  represented by a boolean expression over the
variables $B_1,\ldots,B_k$
\item $S^A$ as the form  $Prog^A=(\mathcal{V}^A,\mathcal{T}^A=\{\tau_1^A,\ldots,
\tau_n^A \}, Init^A)$
\end{itemize}

\paragraph{Abstract transition}
Let $\mathcal{T}^A$, be an abstract transition, it  must  satisfy  the condition of the definition of abstract  program , {\tt s.t.} all  transition $\tau$, $post[\tau^A](P_B) $, where
 $\tau^A $ is the abstract transition corresponding to $\tau$, have to
 represent all  concrete states $q'$ which are  successors by $\tau$ of concrete state $q$ represented by $P_B $.
We must show :: $post[\tau](\gamma(P_B)) \Rightarrow \gamma(P_B^{'})$,

\end{definition}

\subsection{Algorithmic checking of refining}
This model-checking needs methodological and correctness conditions:

\subsubsection{Methodological conditions} 
\begin{itemize}
\item New actions and variables will be introduced by refining
\item the variables of refine system and abstract system must be linked by a ``collage'' invariant. 
\item
\end{itemize}
\subsubsection{Correctness conditions:}
\begin{itemize}
\item simulation of the refine system by the abstract
\item no cycle between the new action
\item no new deadlock
\end{itemize}

\begin{figure}[h]
\includegraphics[scale=0.8]{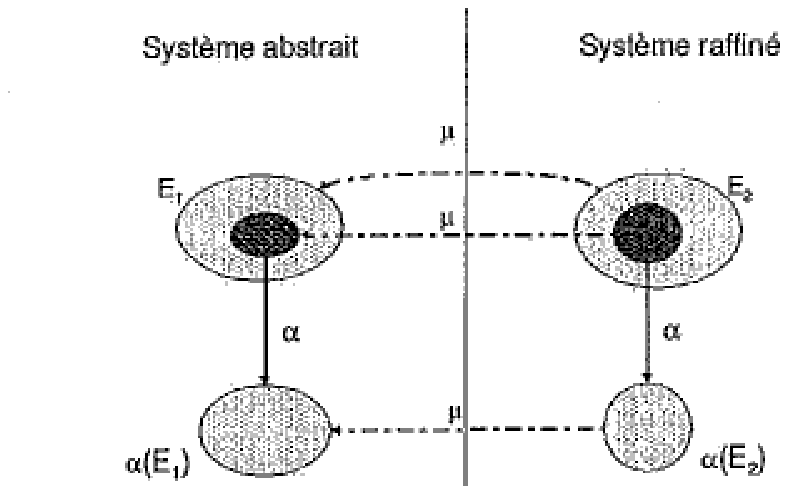}
   \caption{Simulation with old actions}
	\label{fig1}
\end{figure}

\begin{figure}[h]
\includegraphics[scale=0.8]{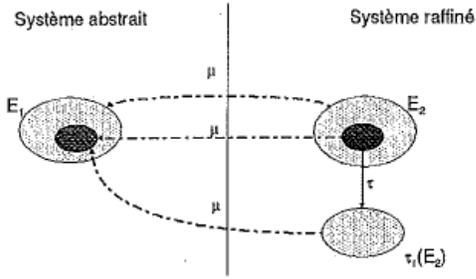}
   \caption{Simulation with new actions}
	\label{fig2}
\end{figure}
It is a question of carrying out an iteratif calculation of the simulation 
of $TS_2$ by $TS_1$ cfr figures \ref{fig1} and  \ref{fig2}  where transisition $\alpha$ is replace by $\tau$.
The algorithm terminates when the fixpoint is reached. 
 
\begin{theorem}
\begin{itemize}
\item If $P_1$ is property satisfied by $TS_1$ and   $TS_2$ refines  $TS_1$ then   
 $$  \frac{TS_1 \models P_1,  \vdash  TS_2 \subseteq TS_1}
       { TS_2 \models P_1} 
 $$
\item If $P_1$ is property satisfied by $TS_1$ and   $TS_2$ refines  $TS_1$ and  if $P_2$ is a reformulation of $P_1$  then 
 $$  \frac{TS_1 \models P_1, \vdash   TS_2 \subseteq TS_1}
       { TS_2 \models P_2} 
 $$ 
\end{itemize}
\end{theorem}

\begin{definition}: Let K, K' two systems (resp concrete and abstract). we call 
false-counterexample or {\em negative-false} a false universal property in K' but true in K 
We say that the counterexample specified in K' cannot be reproduced in K 
\end{definition}

\paragraph{Corollary}
 If K' is too small, it is very probable that it appears 
the negative one. If K' is too large, then the checking is not possible the 
refinement guided by counterexample is thus a natural approach to solve this 
problem by using a adaptive algorithm which gradually creates  an abstraction  
function by the analysis of false-negative: 
\paragraph{ Pseudo Algorithm}
\begin{tabbing}
AA \=  \kill
1. \= {\bf Initialization:} \\
\> generate a first  abstraction function; \\
2. {\bf  Model-Checking:}\\ 
\>  check the model.\\ 
\>  if \=  the checking is a success:\\  
\> \>    then \=\\
\> \> \>     the specification is correct and \\ 
\> \> \>     the algorithm  terminates \\
 \> \>   else  \\
\> \> \>     generate a counterexample from the abstract  model \\
\> \> \>    verify if this counterexample is a negative-false\\
\> \> \>      if It is a success then terminate\\
\> \> \>       else  refine the abstract function such that \\
\> \> \>       the negative-false  can be avoid \\
\> \> \>      goto step 2.
\end{tabbing}
\begin{figure}[h]
\includegraphics[scale=0.8]{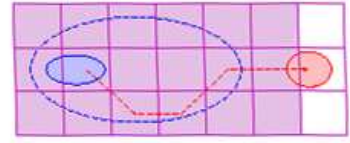}  
   \caption{Example of predicate abstraction}
\end{figure}

\begin{figure}[h]
\includegraphics[scale=0.8]{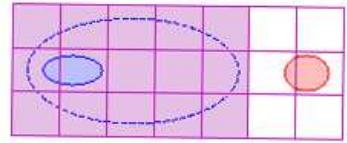}    
   \caption{abstraction too coarse}
\end{figure}

\begin{figure}[h]
\includegraphics[scale=0.8]{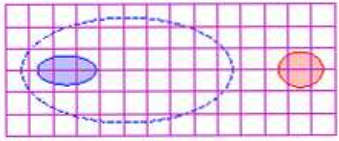}    
   \caption{Example of refinement}
\end{figure}

\section{ Summary}
It is thus a question of starting by carrying out an approximation of a 
way which carries out initial state in a bad condition. Then, a refinement 
"forwards" or "backward" is carried out, and this process is to repeat until a 
fixpoint is met. If the resulting set of states  is empty then the property is 
prove, since one no bad condition is reachable, else, nothing guaranteed the 
value of against example which perhaps distorted by approximation coarse. 
Heuristics is employed to determine the subset of the reachable states since 
the initial states. If an equivalence is found, it really acts of an error 
which can be deferred like a bug, one speaks about positive-false. 
Abstraction by Predicate :
the checking of program by abstraction of closed predicate is a technique of 
checking of program by abstract interpretation where the abstract domain
 is composed of the set  of guard relating to the states and the transitions 
from the system. This domain can be generated automatically and checked by a 
theorems-prover. Like, the set of predicates is always finished, it can be 
coded by a vector of Boolean, which makes it possible on the other hand to use
 the model-checker for calculations of  fixpoint.Si, the domain is very large, 
one can use a  chaotic iterator and to use a widening if it is necessary of 
speed up  the convergence. The termination and reachability decidable in 
this case. The one limitation of this technique of checking by predicate 
abstraction  is that the processes of refinement, which primarily consists in 
calculating the weakest invariant, are extremely slow. 
This obligates the users to require at least the atomic predicate necessary 
to the proof. 
This fact the human intervention which specific is given must be repeated for different programs even if they are very similarities. and it

\section{Conclusion and Future Work}
It has been shown that static checker can cover a large number of potential faults, their automatic usage is still far from realistic. However, as a verification step prior to testing or code review, static checkers, can already enhance the software development process today.
Several techniques like Altarica, B or CSP2B were proposed to specify and
check reactive systems by using hierarchic development by refinement.
 In this case, the systems design is realized gradually by increasing the
systems design to each step of the specification from  a very abstract sight
of the system until its implementation. For us, a system implements (refines)
another system if all the traces of execution of the most detailed system
are too traces of the most abstract ( modulo the introduction of details
during refinement). The checking of the system thus will use refinement to
model the initial system in a more precise way, if the model-checker
provides a erroneous result consequence of  coarse approximation
 at the time of the abstraction.

\bibliographystyle{IEEEtran}
\bibliography{IEEEabrvIEEEexample}

\end{document}